\documentclass[aps,prb,twocolumn,superscriptaddress]{revtex4}
\usepackage{graphicx}
\usepackage{bm}
\usepackage{amsmath}

\begin{document}


\title{
Josephson lattice structure in mesoscopic intrinsic Josephson
junctions by means of flux-flow resistance in
Bi$_2$Sr$_2$CaCu$_2$O$_{8+\delta}$ }



\author{I. Kakeya}
\email[]{kakeya@ims.tsukuba.ac.jp}
\affiliation{Institute of Materials Science, University of
Tsukuba, Tsukuba, Ibaraki 305-8573 Japan}
\author{M. Iwase}
\affiliation{Institute of Materials Science, University of
Tsukuba, Tsukuba, Ibaraki 305-8573 Japan}
\author{T. Yamamoto}
\affiliation{Institute of Materials Science, University of
Tsukuba, Tsukuba, Ibaraki 305-8573 Japan}
\author{K. Kadowaki}
\affiliation{Institute of Materials Science, University of
Tsukuba, Tsukuba, Ibaraki 305-8573 Japan}


\date{October 25, 2004}
\begin{abstract}
Dynamical nature of the Josephson vortex (JV) system in
Bi$_2$Sr$_2$CaCu$_2$O$_{8+\delta}$ (Bi2212) has been investigated
in the presence of the $c$-axis current with magnetic field
alignments very close to the $ab$-plane. As a function of magnetic
fields, the $c$-axis JV flux flow resistance oscillates
periodically in accordance with the proposed JV triangular
structure. We observe that this oscillating period becomes doubled
above a certain field, indicating the structure transition from
triangle to square structure. This transition field becomes lower
in junctions with smaller width perpendicular to the external
field. We interpret that this phenomena as the effect of the edge
deformation of the JV lattice due to surface current of intrinsic
Josephson junctions as pointed by Koshelev.
\end{abstract}


\maketitle


\section{Introduction}
Highly anisotropic high-$T_c$ superconductors such as
Bi$_2$Sr$_2$CaCu$_2$O$_{8+\delta}$ are well described as stacks of
weakly coupled Josephson junctions\cite{Kle94}, where charge
coupling between CuO$_2$ double-layers is not
negligible~\cite{Koy96}. This characteristic feature is
responsible for properties of so-called intrinsic Josephson
junctions (IJJ) like multiple branches in current-voltage
characteristics and longitudinal Josephson plasma
mode~\cite{Mac99}. Appearance of various vortex phases under the
$c$ axis field indeed manifests the weak coupling between the
layers.

In the magnetic field parallel to the $ab$ plane ($\perp c$ axis),
the situation becomes different because of inductive coupling
between Josephson vortices generated by the parallel magnetic
field\cite{Bul94}. In high fields above $\gtrsim$ 1 T, Josephson
vortices occupy all block layers and form the Josephson vortex
lattice which demonstrates the strong interaction between vortex
arrays in the neighboring layers~\cite{Hu_00}. This is the one of
the most striking difference of Josephson vortices in IJJ from
ones in isolated single Josephson junctions. Furthermore, many
characteristic features of Josephson vortex state in parallel
magnetic fields have been found in the $I-V$
characteristics~\cite{Lee95}, the Josephson plasma
resonance~\cite{ISS01}, and the vortex phase diagram~\cite{Mir01}.

Recently, Ooi et al~\cite{Ooi02} found periodic oscillation of the
Josephson vortex flow resistance in small Bi2212 single crystal as
a function of magnetic field. They claimed occurrence of the
coherent flow of the triangular Josephson vortex lattice.
Subsequently, Koshelev~\cite{Kos02} and Machida~\cite{Mac03} have
reproduced the oscillation by calculating the flux flow resistance
by taking the surface barrier effect into consideration. This
means that the surface barrier is not so strong to destroy the
bulk properties of the Josephson vortices, where it prefers to
form the triangular lattice even in the vicinity of sample edges.
This is in contrast to the vortex state in fields parallel to the
$c$ axis where vortex arrangements in the vicinity of sample edges
are dominated by the surface barrier effect.

In this paper, we present experimental results of the Josephson
vortex flow resistance in Bi$_2$Sr$_2$CaCu$_2$O$_{8+\delta}$
micro-structured crystals with dimensions of 1$-$10 $\mu$m. In the
smaller samples and at higher fields, we have found periodic
oscillation of the Josephson vortex flow resistance. This
indicates that the square vortex lattice is realized. This
experimental observation is attributed to the effect of the
surface barrier, which becomes dominant in smaller samples and
overcomes the bulk properties in the short stack regime.

\section{Results and Discussions}

\begin{figure}
\begin{center}
\includegraphics[width=1.0\linewidth]{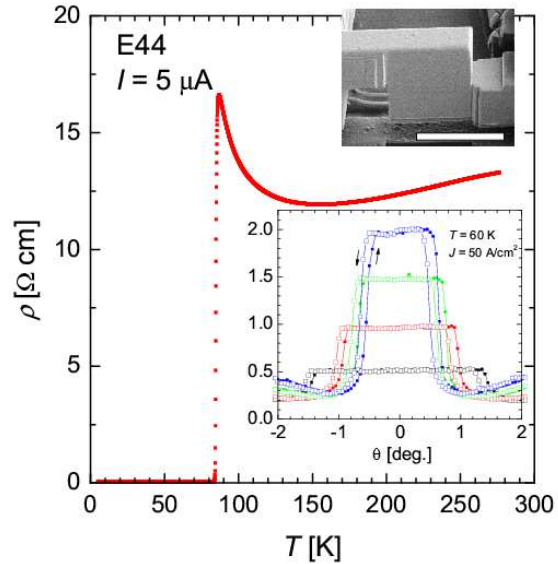}
\end{center}
\caption{ Temperature dependence of the $c$-axis resistivity at
$H=0$ in E44. It is found that the crystal is slightly over-doped
. The lower inset shows angular dependence of the $c$-axis
resistivity of $J=50$ A/cm$^2$ at $H=1$ T in the vicinity of the
$ab$ plane. Arrows indicate directions of change in angle and
slight hysteresis was found. Upper inset depicts SEM image of the
sample. A thick bar corresponds to 10 $\mu$m in the image.
\label{fig1}}
\end{figure}

We made in-line symmetric junctions~\cite{Kim99} to avoid various
contact problems accompanied by the transport measurements of
intrinsic junctions. Bi$_2$Sr$_2$CaCu$_2$O$_{8+\delta}$ single
crystals were grown by the travelling solvent floating zone
method. A cleaved thin ($\sim 10$ $\mu$m) crystal was carefully
cut into narrow strips with a width of $\sim 100$ $\mu$m by a
sharp knife. The narrow strip was put on a MgO substrate with four
gold electrodes which were patterned by a vacuum deposition with a
mask and the center of the strip was shaped into mesoscopic
junctions as shown in the inset of Fig. \ref{fig1} with the
focused ion beam (FIB) machine SMI2050MS (SII NanoTechnology
Inc.). We made junctions from eight different crystals from A to H
and FIB processes were repeated three times in crystal H after
measurements, resulting in eleven samples in total with different
dimensions as listed in Table \ref{tab1}.

\begin{table}
\caption{\label{tab1}List of samples which we measured. Numbers
following capital letters refer to sizes of junctions
perpendicular to the magnetic field and the $c$-axis. }
\begin{ruledtabular}
\begin{tabular}{lllll}
 & $w$ [$\mu$m] & $l$ [$\mu$m] & $t$ [$\mu$m] & $T_c$ [K]\\
\hline
A94 & 9.4 & 10.3 & 0.3  & 86.6 \\
B19 & 1.9 & 11.7 & 0.5  & 90.0 \\
C22 & 2.2 &  8.9 & 0.14 & 83.6 \\
D49 & 4.9 & 17.8 & 0.08 & 88.2 \\
E44 & 4.4 &  9.5 & 0.15 & 84.7 \\
F18 & 1.8 &  9.9 & 1.14 & 87.2 \\
G20 & 2.0 & 10.2 & 0.9  & 84.3 \\
H41 & 4.1 &  5.5 & 2.0 & 86.9 \\
H55 & 5.5 &  4.1 & 2.0 & 86.9 \\
H38 & 3.8 &  5.5 & 2.0 & 86.9 \\
H30 & 3.0 &  5.5 & 2.0 & 86.9 \\
H23 & 2.3 &  5.5 & 2.0 & 86.9 \\
\end{tabular}
\end{ruledtabular}
\end{table}

By applying the $c$ axis current between two current electrodes,
the current is concentrated into small junction area only. The
voltage between the voltage electrodes comes from only the small
junction area because the other parts of the sample still remain
in the superconducting state. The temperature dependence of
resistivity of a junction is shown in Fig. \ref{fig1}, in which a
clear superconducting transition and zero resistance below $T_c$
are obtained. External magnetic fields were applied by the split
pair superconducting magnet which can generates horizontal
magnetic field up to 80 kOe. Rotating magnetic field around the
ab-plane with a precise rotator, the $c$-axis resistance suddenly
increases in the vicinity of the $ab$-plane as shown in the inset
of Fig. \ref{fig1}. This resistance is attributed to the Josephson
vortex flow resistance and the lock-in state without pancake
vortices is realized in this angle range. We can set the field
alignment parallel to the $ab$-plane with an accuracy of 0.01
degree.

\begin{figure}
\begin{center}
\includegraphics[width=1.0\linewidth]{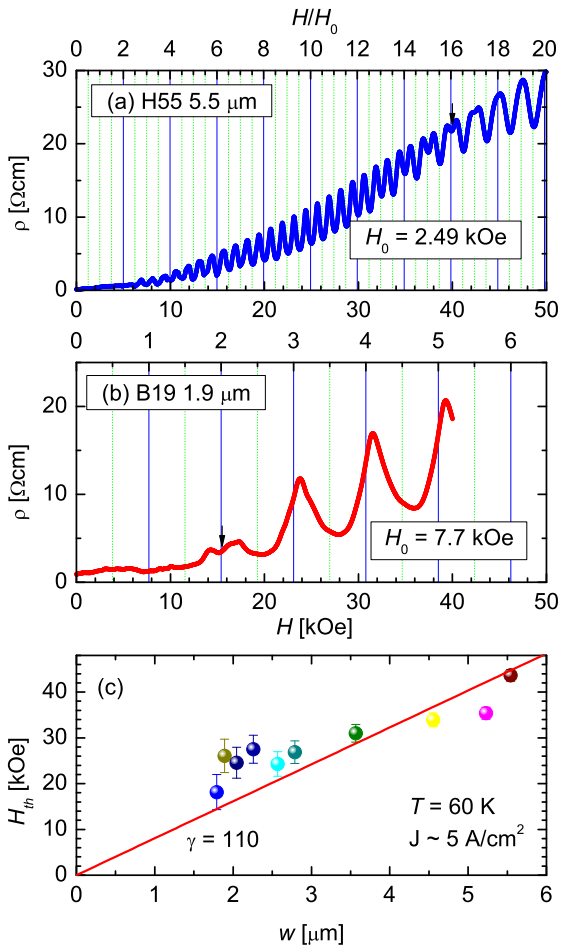}
\end{center}
\caption{Field dependence of the $c$ axis resistivity in H55 (a)
and B19 (b) and threshold field $H_{th}$ of various samples (c).
Arrows in (a) and (b) point respective $H_{th}$ and length of
error bars in (c) corresponds to the period of the $H_0/2$
oscillation of the sample. \label{R-H@H55}}
\end{figure}

Figure \ref{R-H@H55} (a) represents field dependence of the
$c$-axis resistance $R_c$ at 60 K in H55. External field was
applied parallel to the $ab$-plane and swept up to 50 kOe with a
constant $c$ axis current (1 $\mu$A) at a constant temperature.
With increasing field from zero, the resistance smoothly increases
up to 8 kOe, then begins to oscillate with oscillation center
being monotonically increasing. The oscillation period $H_p$ is
constant up to 30 kOe, $H_p =1.25$ kOe, and the amplitude is
maximized around 25 kOe, above which the oscillation gradually
diminishes with increasing field. However above 35 kOe, the other
oscillation in resistivity with different $H_p$ develops and the
oscillation of $H_p =1.25$ kOe seems to disappear above 40 kOe.

$H_p$ below 30 kOe is very close to the field, which corresponds
to add a half flux quantum $\phi_0/2$ to a JV array in a Bi2212
block layer as $H_0/2=\phi_0/2ws=$ 1.25 kOe for H55, where $s=15
\mathrm{\AA}$ is the periodicity of CuO$_2$ double layers of
Bi2212. This oscillation is quite similar to the previous results
observed by Ooi et al.~\cite{Ooi02}, where they argued that this
is due to surface barrier effect for JVs forming the triangular
lattice. The surface barrier for the triangular JV lattice is
considered to be enhanced when the number of JVs in a layer
($H/H_0$) corresponds either an integer or a half odd integer
because the periodicity of the JV lattice matches with the
junction width for both two cases. These two matching result in
the oscillation having period of $H_0/2$ and the bottom at $H/H_0$
being half integers.

\begin{figure}
\begin{center}
\includegraphics[width=1.0\linewidth]{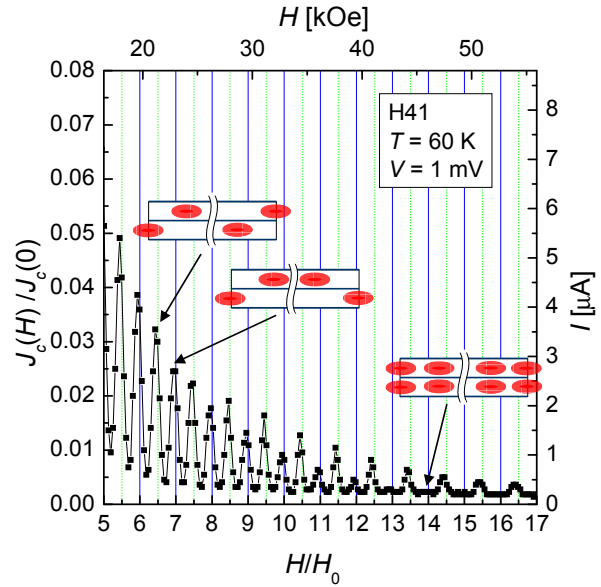}
\end{center}
\caption{Field dependence of $J_c$ extracted from a measurement of
$I-V$ characteristics. \label{Jc-H@H41}}
\end{figure}

$H_p$ above 40 kOe was found to be 2.49 kOe, which is twice as the
period below 30 kOe and very close to the $H_0$ value of 2.50 kOe.
With increasing field in transient region from the $H_0/2$
oscillation to the $H_0$ oscillation, the amplitude of $H_0/2$
oscillation diminishes and the amplitude of the $H_0$ oscillation
develops, resulting in the oscillation amplitude is larger than
one of the $H_0/2$ oscillation. The $H_0$ oscillation starts at
lower fields in samples with smaller width as shown in Fig.
\ref{Jc-H@H41} (b) where $H_0/2$ oscillation is hardly seen even
at the lowest field, while in the largest sample A94 $H_0/2$
oscillation was observed up to the maximum field range. We now
understand that the $H_0/2$ oscillating behavior observed by the
previous measurement by OOi et al. is due to the large sample
width of more than 20 microns~\cite{Ooi02}.

Critical current density $J_c$ as a function of $H_{\parallel}$ of
H41 is shown in Fig. \ref{Jc-H@H41}, which is extracted from $I-V$
characteristics at various $H_{\parallel}$ with a criterion of $V$
= 1 mV. One finds that $J_c(H)$ below 35 kOe oscillates with
period of $H_0/2$ and has a local maxima at $H/H_0$ being half
integers, whereas $J_c(H)$ above 42 kOe shows oscillation with the
period being $H_0$ and local maxima at $H/H_0$ being half odd
integers. Magnetic field dependence of $J_c$ similar to our
results in low fields has been expected theoretically by taking
boundary deformation of JV lattice into
consideration~\cite{Kos02}. Also, $J_c(H)$ in high fields is
similar to magnetic field dependence of $J_c$ in single Josephson
junctions in a sense that the points where the magnetic flux
inside the junction equals to the half odd integer flux quanta
corresponds to the local maxima of the critical current. The
Fraunhofer-like oscillation of $J_c(H)$ is interpreted that the
Josephson vortices form a square lattice where all layers are
identical to edges.

Such a Fraunhofer oscillation in $J_c(H)$ of multilayered system
like Bi2212 has been expected when  the condition $w \ll \lambda_J
\equiv \gamma s$ is satisfied~\cite{Bul92a}. However, $\lambda_J$
is estimated to be in a range from 0.075 to 1.5 $\mu$m for
slightly over-doped Bi2212, which is much smaller than $w$ of this
sample. This discrepancy seems to occur due to inhomogeneity of
the current, which is more realistic in the real specimens of
high-$T_c$ superconductors. It is known that the Bean-Livingstone
surface barrier is pronounced in high-$T_c$ superconductors, so
that vortices prefer to stay at the edge of the sample. This
boundary effect can also explain $H_0/2$ oscillation in $J_c(H)$
as discussed by Koshelev~\cite{Kos02}.

The boundary effect for the JV lattice should be more pronounced
in smaller junctions because the number of vortices at both ends
of vortex arrays inside a layer (usually two), which are most
seriously affected by the boundary effect in smaller junctions has
a greater ratio to the number of total vortices of the array than
in the case of larger samples at constant fields. Consider a
Josephson junction with two JVs as an extreme case. In $\rho-H$
curve as shown in Figs. \ref{R-H@H55} (a) and (b), we define
threshold field $H_{th}$ above which local minima in the vicinity
of $H/H_0$ being an integer cannot be resolved (indicated by
arrows in the figure). It was found that $H_{th}$ is lower in
samples with smaller $w$ from $\rho-H$ curves in samples with
various $w$ as plotted in Fig. \ref{R-H@H55} (c). These results
suggest that the square JV lattice is more favorable for smaller
junctions and higher fields.

Now, let us consider how vortices inside the junctions interact
with edges and each other in transition from triangular lattice to
square lattice. At low fields (although high enough to form dense
lattice), the JV system is dominated by the interaction between
vortices in adjacent layers because the separation between JVs
inside a layer is too far to convey the boundary deformation to
all vortices in the layer. With increasing magnetic field,
vortices inside the layer become closer and closer while the
separation between layers does not change and the interaction
between adjacent vortices in a vortex array and the boundary
effect becomes stronger and stronger because JVs are more closely
packed in the layer. The JV system is finally dominated by the
boundary effect and turns to the square lattice in high field
region at the cost of inductive coupling between layers. The
inductive coupling is weaker in a smaller junction because the
total number of vortices in a layer is fewer, resulting in the
transition from triangle to square at lower fields.

According to theoretical calculation by Koshelev, JVs with their
center being closer to both edges than the lattice deformation
length $l_B=\gamma s (\pi\gamma s^2B/\sqrt{2}\phi_0)$ deviate from
positions extrapolated from the bulk region assuming an undeformed
lattice. Here $\gamma$ and $B$ are the anisotropy parameter and
induced flux density in the sample, respectively. This means
Josephson vortices inside samples with $w < 2l_B$ may not form the
triangular lattice, so that the threshold field above which the
surface barrier effect dominates the system is written as a
function of $w$,
\begin{equation}\label{Bth}
H_{th}=\frac{w}{\gamma s}\frac{\phi_0}{2\pi \gamma s^2}.
\end{equation}
A thick line in Fig. \ref{R-H@H55} (c) corresponds to Eq.
(\ref{Bth}) with $\gamma=110$. The agreement is rather good and
the value of the obtained $\gamma$ is reasonable for slightly
over-doped Bi2212.

\section{Conclusion}
We have investigated Josephson vortex lattice structure in
mesoscopic ($\sim\mu$m) intrinsic junctions with the probe of the
JV flow resistance along the $c$ axis. For the small ($w < 6$
$\mu$m) mesas, the JV lattice is found to change the structure
from triangular to square lattice with increasing fields in low
current limits. Transition field from the triangular to square
lattice monotonically increases with $w$. This is attributed to
the interplay between the surface barrier effect and the inductive
coupling between Josephson vortex arrays in intrinsic Josephson
junctions.

\begin{acknowledgments}
We thank M. Machida and A. Koshelev for
their fruitful discussions and critical comments. This work has
been partially supported by the Ministry of Education, Science,
Sports and Culture, Grant-in-Aid for Young Scientists (B),
14740201, 2002-2003, and 21st Century Center of Excellence (COE)
Program, ``Promotion of Creative Interdisciplinary Materials
Science for Novel Functions''.
\end{acknowledgments}

\bibliography{MyBibs0601,MyNewPubs}

\newpage

\end{document}